\newcommand{\CLASSINPUTtoptextmargin}{19mm}
\newcommand{\CLASSINPUTbottomtextmargin}{25.4mm}
\newcommand{\CLASSINPUTinnersidemargin}{16mm}
\newcommand{\CLASSINPUToutersidemargin}{16mm}

\documentclass[conference,10pt,a4paper]{IEEEtran}
\IEEEoverridecommandlockouts
\usepackage{cite}
\usepackage{amsmath,amssymb,amsfonts}
\usepackage{algorithmic}
\usepackage{graphicx}
\usepackage{textcomp}
\usepackage{xcolor}
\def\BibTeX{{\rm B\kern-.05em{\sc i\kern-.025em b}\kern-.08em
    T\kern-.1667em\lower.7ex\hbox{E}\kern-.125emX}} 

\renewcommand\IEEEkeywordsname{Keywords} 

\newcommand{\IEEEtitle}[1]{\title{\vspace{-6.5mm}#1}}

\usepackage{cite}
\begin{document}
\IEEEtitle{Learning Joint Embedding for Cross-Modal Retrieval}
\author{
\IEEEauthorblockN{Donghuo Zeng}
\IEEEauthorblockA{\textit{National Institute of Informatics, SOKENDAI, JAPAN}}
}
\maketitle

\begin{abstract}
A cross-modal retrieval process is to use a query in one modality to obtain relevant data in another modality. The challenging issue of  cross-modal retrieval lies in bridging the heterogeneous gap for similarity computation, which has been broadly discussed in image-text, audio-text, and video-text cross-modal multimedia data mining and retrieval. However, the gap in temporal structures of different data modalities is not well addressed due to the lack of alignment relationship between temporal cross-modal structures. Our research focuses on learning the correlation between different modalities for the task of cross-modal retrieval. We have proposed an architecture: Supervised-Deep Canonical Correlation Analysis (S-DCCA), for cross-modal retrieval. In this forum paper, we will talk about how to exploit triplet neural networks (TNN) to enhance the correlation learning for cross-modal retrieval. The experimental result shows the proposed TNN-based supervised correlation learning architecture can get the best result when the data representation extracted by supervised learning.
\end{abstract}


\section{INTRODUCTION}
With the improvement of innovative technology and social media, multimedia data and information have explosively emerged on the Internet. The heterogeneous gap of different data modalities causes difficult for learning correlation among the different data modalities~\cite{automaticmusic}. Previous research tried to bridge the gap by formulating a joint embedding space, which has succeeded in the area of cross-modal retrieval. However, learning the cross-modal correlation of static data is not enough for the multimedia mining and retrieval system. Exploring the correlation between temporal cross-modal multimedia data also is a very important role. This motivates us to develop architectures to learn joint embedding for cross-modal retrieval between different temporal modality data.

\textbf{Data representation} is a main challenge of this study, which means how to learn and represent different features from different modality data. Current feature extractors for audio data such as SoundNet\footnote{https://github.com/cvondrick/soundnet} and VGGish\footnote{https://github.com/tensorflow/models/tree/master/research/audioset}, which are wildly applied in sound recognition~\cite{tokozume2018learning} and audio event understanding~\cite{gemmeke2017audio}. 
Visual feature can be extracted by Inception\footnote{https://github.com/google/youtube-8m/tree/master/feature\_extractor} and I3D\footnote{https://github.com/deepmind/kinetics-i3d} model, which have succeeded in computer vision related work~\cite{miech2017learnable}. 

\textbf{Cross-modal correlation learning} is another main challenge for this study, which is to map the different modality data representations into a common space by maximizing the correlation. A typical solution to utilize canonical-correlation analysis (CCA)~\cite{hardoon2004canonical}, which uses linear projections to learn
the correlations between two different modalities based on
SVD. DCCA~\cite{audiolyrics, andrew2013deep} provides complex nonlinear features transformations
of different modalities of data through deep
neural networks. CCA and DCCA will fail to obtain label
information because they only focus on pairwise correlation
learning. Cluster-CCA~\cite{rasiwasia2014cluster} establishes possible one-to-one correspondences to contain the category information during training stage. The category-based DCCA (C-DCCA)~\cite{yu2018category} not only considers the instance-based correlation but also learns the category-based correlation. 

In this work, taking audio-visual as paired sequential dataset samples, we suggest to train a supervised architecture based on the idea of taking CCA as feature extraction to embed audio and visual into the joint space, followed by triplet networks to take negative samples into account for training stage. Some experimental results run over the large-scale audio-visual datasets VEGAS~\cite{zhou2018visual} show that our method outperforms the existing state-of-the-art works, The proposed method also can be exploited in  other paired multimedia dataset.
\begin{figure}[htbp]
\centering{\includegraphics[width=7.8cm, height=8.3cm]{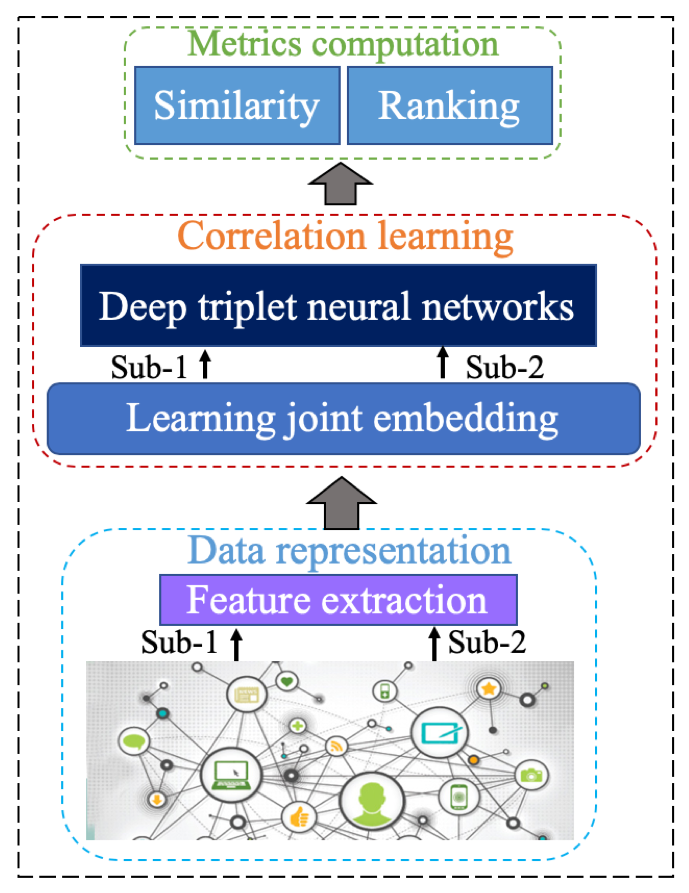}}
\caption{The overview of TNN-based supervised learning architecture. The training process can be described from three different parts. (1) data representation: the raw audio/visual data are represented by some state-of-the-art feature extractors. (2) correlation learning: the audio/visual feature representations are fed into a common space to learn the CCA-based joint embedding, the paired embedding has capability of boosting its internal correlation by TNN model. (3) Metrics computation: the similarity matrix of audio and visual is computed by the generated deep audio and visual representations from TNN model.}
\label{mainmodel}
\end{figure}
\section{APPROACHES}
\subsection{Problem formulation}
As for video $i$, it consists of audio track $a_{i}\in{R^{128\times L \times M}}$ and visual track $v_{i}\in{R^{1024\times L \times M}}$, where L is the number of extracted frames for video $i$, M is the number of videos. Take audio-to-visual as an example, when we input audio as query, the system will predict a ranking list of videos based on the correlation between the query and all visuals.
\begin{gather*} 
Ranklist(a_{i}) = sortcorr_{j\in [1,M]}(a_{i}, v_{j})
\end{gather*}

Where the $corr(\cdot)$ function is computed for the pairwise correlation and similar-dissimilar correlation by TNN architecture.

\subsection{Architecture of our approach}
Fig.~\ref{mainmodel} shows our TNN architecture consists of two different branches. The raw data of each branch is extracted by an advanced time-sequence pre-trained model, the raw data is represented by a vector per second. In our architecture, we add a mean layer for each branch before correlation learning. In the training process, we construct one-to-one possible correspondences as the training set from data representation of audio-visual based on their concepts, then feed training set into a common space to learn joint embedding by maximizing the correlation between them. However, for each query of one modality, CCA-based methods only calculate correlation from similar data of other modality. Triplet-based networks computes similar and dissimilar data of other modality by pushing these dissimilar data and pulling these similar data for the query, which can enhance this correlation in the same space and generate a better embedding for each branch. Based on the generated embedding, they can be used for similarity matrix computation. Once The similarity matrix is generated, it is not hard to obtain a ranking list for each query to evaluate the performance of our architecture.

\section{PRELIMINARY RESULTS}
Our proposed architecture~\cite{zeng2018audio,ccatriplet2019} is evaluated on the audio-visual dataset: VEGAS~\cite{zhou2018visual}. Each audio-visual pair belongs to a single concept and the audio-visual data is represented by high-level features. We adopt the task of cross-modal retrieval to conduct the experiments and assess the result with mean average precision (map). In our experiment, the dimensions of audio and visual are set to 128*L and 1024*L, where the L is the number of extracted frames for audio-visual. The dimension of output from architecture for each audio/visual is 10 respectively. The experiment result achieved by S-DCCA with different data representations is shown in table~\ref{result}. It is noted that supervised-based data representation VGGish/Inception can achieve the best results. Table~\ref{result2} shows that the TNN model outperforms S-DCCA and other approaches based on VGGish/Inception features. This demonstrates that TNN model can enhance the correlation between audio and visual.

\begin{table}
\centering
    \caption{The map of S-DCCA with different data representation.}
    \label{result}
    \begin{tabular}{c|c|c} 
    \hline 
       Features (audio/visual) &audio-visual  &visual-audio\\  \hline
       SoundNet/I3D        &25.45\%   &25.28\% \\
       VGGish/I3D          &47.49\%   &46.22\% \\
       SoundNet/Inception  &49.82\%   &48.63\%\\
       VGGish/Inception    &70.34 \%   &69.27\%\\
      \hline
    \end{tabular}
\end{table}

 \begin{table}
\centering
    \caption{Comparisons of the cross-modal retrieval results on VEGAS dataset with the state-of-the-art methods }
    \label{result2}
    \begin{tabular}{c|c|c} 
    \hline 
       Models &audio-visual  &visual-audio\\  \hline
       CCA     &32.43\%   &32.11\% \\
       DCCA    &14.09\%   &12.49\% \\
       S-DCCA  &70.34\%   &69.27\% \\
       \textbf{TNN}     &72.69\%   &71.84\% \\
      \hline
    \end{tabular}
\end{table}
\section{WORK IN PROGRESS}
In our previous research, learning joint embedding for cross-modal retrieval remains two main challenges. The aim of effective and reliable data representation is to improve the accuracy of multimedia mining and retrieval system. Meanwhile, the architecture can be learned accurately by the correlation between audio/visual on data representations.

The proposed TNN architecture is not only used for temporal structure audio/visual data but also is suitable for other multimedia data such as static structure image-text data, temporal and static structure based video-text data. In the future, we will evaluate this architecture for other multimedia data, and propose more advanced architecture.

Recently, an advanced deep learning algorithm generative adversarial network (GAN) has been introduced into multimedia data mining and cross-modal retrieval area, where it has proved GAN-based methods can surpass traditional methods. In the following time, we plan to apply a generative model to generate more effective features for multimedia data under the constraint of the discriminative model by adversarial learning. The discriminative model consists of a deep triplet neural network which can better distinguish the original data and generated data.
\bibliographystyle{IEEEtran}
\bibliography{mybibliolist}
\end{document}